\documentstyle[prb,aps,multicol,tighten]{revtex}

\begin{document}

\title{Inertia in the Brazil Nut Problem}
\author{Y. Nahmad-Molinari, G. Canul-Chay and J. C. Ruiz-Su\'arez \cite{JCRS} \\
Departamento de F\'{\i}sica Aplicada, CINVESTAV-IPN, Unidad\\
M\'erida, A. P. 73 ``Cordemex,'' \\
M\'erida, Yucat\'an 97310, M\'exico.}
\date{\today}
\maketitle

\begin{abstract}
The rise dynamics of a large particle, in a granular bed under vertical
vibrations, is experimentally studied with an inductive device  designed
to track the particle while it climbs through the granulate  under different
conditions. A model based on energy considerations is presented to explain
our experimental data, drawing the important conclusion that it is the
inertia of the particle, assisted by Reynolds dilatancy,
the driven force behind its ascension mechanism. The ascension reveals a
friction profile within the column which remains unchanged for different
accelerations.
\end{abstract}


\begin{multicols}{2}


Industries dealing  with non-consolidated granular
materials are aware of the following phenomenon: large particles in a bed of
small ones rise and segregate to the top when the system is vertically
shaken. This is an important issue, since very often granular mixtures loose
their homogeneity properties when they are in vibrating environments (see,
for instance, a recent review by A. Rosato \cite{Rosato 2002}). When particle  
separation in granular beds  is the aim of an industrial process,
the phenomenon can be used in our favor. In any case, a complete
understanding of this problem is a scientific challenge not completely
fulfilled despite the great number of experimental and theoretical studies
carried out in the last years \cite{Muzzio,Vanel,Herrmann,Rosato 1987,Nagel
93}.

One of the first quantitative attempts to understand granular segregation using
dynamic simulations, was reported in 1987 \cite{Rosato 1987}. The problem,
baptized by Rosato and coauthors with the appealing name of ``Brazil Nut
Problem'' (BNP), has been theoretically studied ever since, using several
computer models \cite{Herrmann,libro1,libro2,Jullien}. Such studies focused
mainly on the influence of size, friction and excitation parameters. Some of
these results support the hypothesis that it is reorganization or
``void-filling'' beneath large particles, the universal mechanism promoting
their  upward movement through the bed. However, 
experimental studies find evidence that it could be global convection rather
than reorganization, the driven force behind the BNP \cite{Nagel 93}. The
dilemma is not yet settled as one can see in very recent reports 
\cite{Rosato 2002,Nagel 2001}. M$\ddot{o}$bius et al carried out an 
interesting experiment in which large
intruders, of different densities, rise to the top of a vibrating granular
column \cite{Nagel 2001}. Two important aspects they observe are the 
following: air is
important in the transport of the intruders through the bed, and decreasing
the density of the intruder does not necessarily mean a monotonic increasing of
the rise time, as might be suggested by studies in 3D \cite{Muzzio} and 2D 
\cite{Liffman}. A new experimental work studying the influence of air drag
in granular  segregation has recently appeared \cite{Burtally},  in which suppression
of the segregation phenomenon has been observed if air is completely
removed from  fine binary glass-bronze mixtures of equally sized spheres (90 to
120$\mu m$).

The aim of this manuscript is to report an experimental study that might
contribute to the understanding of the physical mechanisms behind the
BNP. With the exemption of few experimental studies \cite{Nagel 93,Liffman},
the experiments devised so far to study the BNP have been centered in
measuring rise times of an intruder that climbs through a vibrated granular column. 
These measurements showed to be important to describe the
physics of the problem, but as we will see below they do not give a
complete picture of it. 

A new magnetic technique
is proposed here to track the movement of a metallic bead in a granular column under
vertical vibrations. Not only rise times are measured, but the complete
dynamics of the bead is obtained.

In Fig.\ref{setup} a schematic of the experimental set up is shown. A
cylindrical column is made of Plexiglas, with inner diameter of 2.5 cm and
20 cm of length. The Plexiglas tube is fixed to a vibrating table. This table
is fed with an amplified periodic voltage coming from a function generator
(HP-33120A). On the outer surface of a wider Plexiglas cylinder tube, a
solenoid is made with a non-uniform number of turns per unit length. The
first segment of the solenoid is formed by two loops, an empty space of 8 mm
is left and a segment of 4 loops follows. A third segment is made with  6
turns and is also separated by the same distance from the previous one. The
array continues in this fashion until the last segment of the solenoid has
28 turns. From the first segment to the last, there is a total length of 21
cm. The separation between the end of each segment to the beginning of the
next one, is always 8mm. This non-uniform cylindrical solenoid is made with a
thin copper wire. Finally, the solenoid is held from above concentrically to
the granular column, without touching both, the vibrating table and the column.

In a typical experiment the column is filled with small seeds or glass
beads. Before pouring the beads into the column, a large sphere is put
at the bottom. If there is some magnetic contrast between the bead
and the granulate  \cite{note1}, the inductance of the solenoid will
change as a function of the bead position h. The inductance L of the
solenoid is measured by a HP4284 LCR-meter at a frequency of 10 KHz.  What we
measure is the  effective value of L, but L changes according to where
the particle is in a given moment (for instance, at the beginning of the 
experiment the sphere is at the center of the first
solenoid segment and therefore, the inductance has its lowest value). 
A previous calibration of the instrument
is manually done by measuring L as a function of ${\it h}$ (obtaining after
a polynomial fit an almost linear smooth function $L=f(h)$). Thereafter, the
experimental data L vs $t$ obtained in a run are converted into ${\it h}$ vs
$t$. With the device and procedure just described we are able to continuously
follow the ascent of the large particle in the granulate.

Our first experiment measured the dynamics of a spherical steel bead in a
granular column of nearly monodisperse cabbage seeds as a function of the
vibrational amplitude. The diameters of the bead and the small particles
are, respectively, 6.32 and 2 mm (their densities are, respectively, 7.8 and
1.1 g/cm$^{3})$. We maintain fixed the frequency of the vibration at 7.5
Hz. The amplitudes are measured optically using a laser diode, a flat mirror
mounted on the vibration table and a screen one meter away from the table.
Before each run, the metal bead is held at the bottom of the column
magnetically while the system is shaken and compaction of the bed obtained.
Once the column compacts to an equilibrium height, the ball bearing is
released and the experiment begins. Normally, the granular column becomes
electrically charged after several runs, thus, in order to avoid this problem,
each run is made with new and fresh seeds. Excellent reproducibility of the
data is obtained as long as temperature and humidity do not change during
the experiments.

Six experimental curves of h versus number of cycles ($t\omega /2\pi $), for
different vibration amplitudes (frequency fixed), are shown in Fig.\ref{results1}a. As we can
notice, the dynamics of the bead is strongly affected by the vibration
amplitude. Once the bead has climbed the entire column, one 
can see that a little oscillatory re-entrance occurs (see data points
around 11 cm of height). The aspect ratio of the column (height divided by width) is important in order 
to produce a more pronounced re-entrance (or whale \cite{Muzzio}) effect. This 
point will be discussed later.

A quantitative interpretation of the experimental results shown in  Fig.\ref{results1}a
certainly requires a theoretical model. Let us first propose  the following physical picture 
of what we believe occurs during the
ascension of the bead in the granulate: we define time equal zero the time when the
entire system (bottom plate, walls, bead and granulate) has, in its sinusoidal motion, a maximum
upward velocity ($A \omega$) and  zero acceleration.
Time runs and the system starts to decrease the velocity  due to its increasing negative 
acceleration. Before the acceleration of the system reaches the value of $-g$, the column
behaves as a normal silo resting on the ground \cite{Janssen}.
However, just when $a = -g$ the bead and granulate loose contact with the bottom plate of
the cell. Due to wall friction,  inter-particle interactions and the relative acceleration between the
cell and the granulate, stresses reorganize and arches invert, from silo
like shapes (inverted V's) to V shapes. Such phenomenon has been observed
previously in 2D vibrating piles  \cite{Duran93,Duran94,Duran96}. Moreover, due to
this instantaneous stress reorganization or Reynolds dilatancy,  the granular bed soon adjusts to the
same acceleration of the  walls (wall friction is always present and  shear
the  bed, dragging it downwards).  If the walls of the cell had no  friction, once the negative 
acceleration of the cell  reached  and surpassed the value
of $-g$, the entire granulate would loose contact with 
the bottom and travel freely as particles thrown upwards. 
In this condition,  there would  not exist a mechanism to delay or stop most of the bed 
particles and the intruder can not climb through the 
column. 

Although the intruder  could feel the granular stress reorganization, due to its larger 
kinetic energy still follows a ballistic uprise, penetrating by inertia into the bed  a small distance 
that we will  call, penetration length ($P_l$).
Inasmuch as we are
plotting in Fig.\ref{results1}a h as a function of number of cycles (not
time), it is clear that the derivative of any of these curves is  precisely $P_l$ (not
velocity). 

The above physical picture suggests that on each cycle, the kinetic energy
of the bead is lost by friction during its  penetration into the granular bed. 
Therefore, a simple energy balance per cycle gives the following relationship:

\begin{equation}
1/2 m v_{to}^2 = \beta(h) P_l,  
\label{pl}
\end{equation}
where $v_{to}$  is 
the "take-off" velocity the bead has when the  system reaches a negative  
acceleration  $a = -g$,  $m$ the mass of the bead, and $\beta(h)$ the friction force exerted upon 
it by the granulate. $v_{to}$ is simply the value of $\dot z(t)$ when $\ddot z(t) = -g$ (where $z(t) = 
Asin(\omega t))$. Thus,

\begin{equation}
v_{to}= [A^2 \omega ^2 - g ^2/\omega ^2]^{1/2}  
\end{equation}

Eq. \ref{pl} implies that most of the kinetic energy of the bead per cycle is dissipated by friction
(indeed, the potential energy per cycle $mgP_l$ is negligible).
Eq.\ref{pl} can be tested by plotting $P_{l}$  as a function of $A$. 
Clearly, the parabolic behavior predicted by Eq.\ref{pl} is obtained, see Fig.\ref{results2}. 

This parabolic behavior holds at any height $h$ but in order to illustrate it more clearly we use the 
slopes of the lines shown in the inset of Fig.\ref{results2}, which are linear fits to the
lower parts of the ascension curves in Fig.\ref{results1}a. 

Furthermore, since the left term of Eq.\ref{pl} is a constant depending only
on the parameters of the vibration (does not depend on $h$), the right term must be
a constant as well, and thus,  $\beta(h)$ must be a function 
inversely proportional to $P_{l}$. In other words, the friction $\beta(h)$ the bead 
encounters along the column can be obtained
directly by taking the inverse of $P_{l}$. Fig.\ref{results1}b
shows $P_{l}/A^{2}$ as a function of $h$ for amplitudes 1.25, 1.20, and 1.15
cm. Surprisingly, the curves collapse into one, indicating that conservation of energy
given by Eq.\ref{pl} might be considered the correct mechanism behind the BNP. 
This collapse also implies  that  the parabolic behavior of $P_l$ vs $A$ occurs 
at any $h$.

The three other curves, for $A$ equal to
1.30, 1.40, and 1.50 cm, are not shown here for the sake of clarity (they
are increasingly noisy due to the smaller number of experimental points and the numerical 
derivative process, but behaves the same way as the others as could be inferred by 
looking at Fig.\ref{results2}). 
One can see by looking at Fig.\ref{results1}b that the intruder feels the
same friction profile during its ascension through the column at least 
for our acceleration conditions ($\Gamma = (2.5;3.4)$).

Eq.\ref{pl} can be further tested in the following way: we investigate the 
ascension dynamics
of a bead as a function of its own diameter. Four different bead diameters
are considered: 4.74, 6.32, 9.46 and 11.10 mm. In each run, the acceleration
of the vibration table remains fixed; the frequency being again 7.5 Hz and
the amplitude of the vibrations 1.3 cm. The four  beads have the same density as
in the previous runs and the granular column the same height, 12 cm.
Experimental curves of the ascension dynamics of the beads are not shown
since they are similar to the ones presented in Fig.\ref{results1}. However,
plotting $P_{l}$  (evaluating the slopes the same way as in the inset of 
Fig.\ref{results2}) as a function of bead diameter,  we 
obtain the straight line shown in Fig.\ref{results3}. This outcome is the expected
behavior predicted by Eq. \ref{pl}; being $\beta $ proportional to the cross section
of the bead $D^{2}$
and the mass of the particle to $D^{3}$, one obtains $P_{l}\propto D$. 
As in the previous case, this linear behavior holds at any $h$.

If the underlying mechanism behind the BNP  is related to 
stress-chain formation, which in turn is affected principally
by the roughness of the internal walls and by their separation, then the
ascension dynamics of the bead must change  with  $\sigma $ (cell diameter) and $\mu$ (wall friction 
coefficient). Fig.\ref{results4} shows three experimental 
curves (A, B and
C) for the ascension dynamics of a bead in three granular columns with
different diameters (5.3, 4.4, and 2.5 cm) under the same excitation
conditions (frequency of 7.7 Hz and amplitude of 1.2 cm). Clearly, the bead
climbs faster through the granulate the wider is the container. In fact, arches are
weaker (they span more distance) and therefore more easily disrupted by the
ascending bead. 
A turning point would be the case when the separation of the walls 
is much larger. If they go to infinity (very large $\sigma$) 
Reynolds dilatancy on the bed caused by wall friction would become  unimportant. 
The bed, together with the
intruder, would move up and down imponderably and the intruder can not ascend. 
To observe this
condition with our experimental set-up is not possible but is in agreement
with the reported  behavior in fine granular binary mixtures
in vacuum by Burtally \cite{Burtally} and coworkers, in which the walls are
hundreds of bead diameters away from each other. Indeed, they observe that air drives 
segregation and vacuum promotes mixing. In other words, air in this  experiment
is the mechanism to delay the lighter (kinetically poorer) phase of the mixture and  
segregation is seen. On the other hand,  when air is evacuated,
the kinetic energy contrast between the glass and brass spheres is no longer relevant and
convection dominates, mixing both species.
 
Curve D in Fig.\ref{results4} is obtained when the internal
walls of container A are covered by sandpaper. The higher the wall friction coefficient, the
stronger the arch formation effect and therefore the larger the rise 
time of the bead. This effect is surprisingly large if we compare curves D and A.

We now come back to the discussion of the re-entrance effect previously mentioned.
As we said before, a very small (less than 1 cm) re-entrance into the granulate is observed 
in some curves of Fig.\ref{results1}a. We also called the attention to the fact that the aspect ratio
of the column is important to reduce or increase this whale effect. Thus, in order to
study the re-entrance of the intruder we use a wider column keeping the same height for the granulate (12 
cm). Instead of 2.5 cm (as in Fig.\ref{results1}), we use a 5.3 cm diameter column. The ascension curve of 
the intruder in this
case is curve A shown  in Fig. \ref{results4}. Data taken beyond the rise time of the
intruder (not shown in curve A in Fig. \ref{results4}) are depicted  in Fig. \ref{results5}.
An oscillatory behavior is evident with a re-entrance of about 6 cm. The intruder 
performs an oscillatory movement within a large convective cell formed
in the upper part of the container as it may be expected. The
formation of these convection cells has been attributed to wall friction
and a granular temperature gradient along the column and has been studied by
NMR, dyed tracer particles, molecular dynamics simulations and image
velocimetry. In some papers, the BNP is attributed to this convective motion 
\cite{Nagel 93,Nagel 2001}. As it can be seen in Fig. \ref{results5}, the convection cell
is strong enough to force the bead on top into the granulate again. 
However, since in our experiments we never observed an oscillatory movement with amplitude of 12cm
(the height of the bed), either the convection cell does not span the entire granulate, or it does
but is not able to drag the intruder through the entire column. 
Hence, in our conditions (where the intruder is positioned at the 
bottom of the granulate) convection can not be the mechanism behind the BNP. 

To confirm this,  some convection experiments were carried out. 
We prepared tracer particles  (black and
 red), put the  red ones at the bottom of the column (a 
cylinder with
 inner diameter of 5.3 cm) and  fill it  with  
 particles of the same size and density (the height of the granulate was
 12 cm). We put afterwards the black tracer particles on top
 of the granular bed.
 First,  we study  the convection with no intruder (to avoid any 
 disturbance of  the possible convection cells). We tried different combinations of frequencies and 
amplitudes within  the  accelerations values reported above  and
 the rise  times of the  red particles were carefully measured. In all our 
 experiments the red  particles diffused to the top very slowly,
 in no less than 20 minutes.
 The black particles went into shallow convective cells, down to 3 up
 to 5 cm below the surface. Descending  times  were smaller  but still
 large (few minutes).
 We did now, and separately,  the experiment with different intruders
(metal beads as above). The ascension times from the bottom to 
 the top of the intruders were, for our accelerations, in any case no 
 longer than one minute. No matter the conditions we tried, the intruder 
 climbed  much faster than the tracer particles.  We conclude that, in our 
conditions, the ascension of the 
 intruder is not correlated  with the convective cells. Indeed, in 
 some experiments the intruders climbed through the 
 granulate in 5 or 6 seconds, whereas tracer
 particles did it in 20 minutes at the same conditions.

Although  full convection is not observed in narrow  
columns, one might 
think that a crossover  between inertia and convection takes place for wider 
cells, where granular dragging  in the upper part of the column  
could enhance  the ascension velocity of an intruder.

In a very recent experiment  \cite{Nagel 2001}, where only rise 
times of a bead climbing a granular bed are measured, the bead is positioned not in the 
bottom of the granulate, but very close to the top of the bed where a convective cell certainly exists.  
In that experiment it is seen  that the rise time of the intruder is a non-monotonic function of the 
density. This non-monotonic behavior  could be explained by the competition of two mechanisms: 
inertia and convection. For low intruder densities, convection seems to be  dominant and the rise 
time drops when  the intruder density decreases (the lighter the intruder is, the more easily dragged to the 
top). Inertia, however,  becomes important when the intruder density 
increases and rise times
drops again as inferred by our model. Here, it is clear that air 
acts as a lubricant. When the cell is evacuated, there is no lubricant
and the drag caused by the convective flux is  strong enough to carry  the intruder upwards at the same 
speed regardless its weight, observing the same rise time \cite{Nagel 2001}. But,  if the intruder starts
its ascension from the bottom of the column (where there is no convection), the rise time of the intruder
does depend on its density whether or not air is present.
Very recently, it has been experimentally established in 2D that rise times monotonically decrease with the
mass of the intruder \cite{Liffman}. Our results agree with those 2D  observations.
Recently, we became aware of new results concerning the BNP  
that confirm also our findings \cite{chinos}.

We present experimental results that shed new light on the
fascinating BNP. All previous experimental work carried out on this problem
has mainly focused on measuring rise times, without measuring the complete
non-linear dynamics of the rising particle. Our gradient-coil technique
solves this problem. Based on our data and a simple kinetic energy argument
we conclude that in the BNP the bead climbs the granular column driven by
inertia, assisted by  Reynolds dilatancy through stress-chain 
formation. We used
the velocity as the main parameter and not the dimensionless
acceleration $\Gamma $ in order to stress out the role of the bead-bed kinetic energy
contrast in the climbing mechanism.


YNM and GCCH acknowledge, respectively, scholarships for graduate and
undergraduate studies from Conacyt, and Cinvestav, M\'exico. This work has
been partially supported by Conacyt, M\'exico, under Grant No. 36256E.


\begin{figure}
\caption{A schematic of the experimental setup. A solenoid with a
non-uniform number of turns per unit length is held concentrically to a
vibrating column full of small particles. A larger metallic particle inside
the column drifts up due to the BNP effect. At each vertical position, the
climbing bead feels the magnetic field gradient inside the solenoid,
changing accordingly the value of the solenoid inductance.}
\label{setup}
\end{figure}

\begin{figure}
\caption{A) Six experimental curves of h vs oscillation cycles ($\omega t/{%
2\pi}$). Each curve represents the ascension dynamics of a bead in a
granular column vibrated at different amplitudes: 1.15, 1.20, 1.25, 1.30,
1.40, and 1.50 cm. B) The derivative of curves corresponding to $A$ = 1.15,
1.20, and 1.25 cm divided by the square of $A$. The curves collapse into
one, where the solid line is only a guide to the eye. The dashed line is the inverse of the solid 
one and is proportional to the friction force.}
\label{results1}
\end{figure}

\begin{figure}
\caption{$P_l$ as a function of vibrational amplitude A. The values of $P_l$
in this figure are the slopes of the lines shown in the inset, which represents
the lower parts of the ascension curves shown in Fig.\ref{results1}a. The solid line is
a parabolic fit put in the figure only as guide to the eye.}
\label{results2}
\end{figure}

\begin{figure}
\caption{Penetration length $P_l$ as a function of bead diameter.}
\label{results3}
\end{figure}

\begin{figure}
\caption{Experimental curves (A, B, and C) representing the ascension
dynamics of a bead in granular columns with different diameters: 5.3, 4.4,
and 2.5 cm. Arrows indicate the rise time of the bead. Curve D was obtained
in a run using the column with the largest diameter (5.3 cm), but with a much
greater wall friction coefficient}
\label{results4}
\end{figure}

\begin{figure}
\caption{Oscillatory re-entrance of the intruder into the granular bed caused by a convective cell on 
the top of the granulate for the widest cell.}
\label{results5}
\end{figure}

\end{multicols}


\begin{thebibliography}{*}
\bibitem{note1}  Even if the magnetic susceptibility contrast of the bead
and bed particles is not large (both could be diamagnetic) our LCR-Meter is
sensible enough to measure a change in L when the bead moves in the bed
column.
\end{thebibliography}
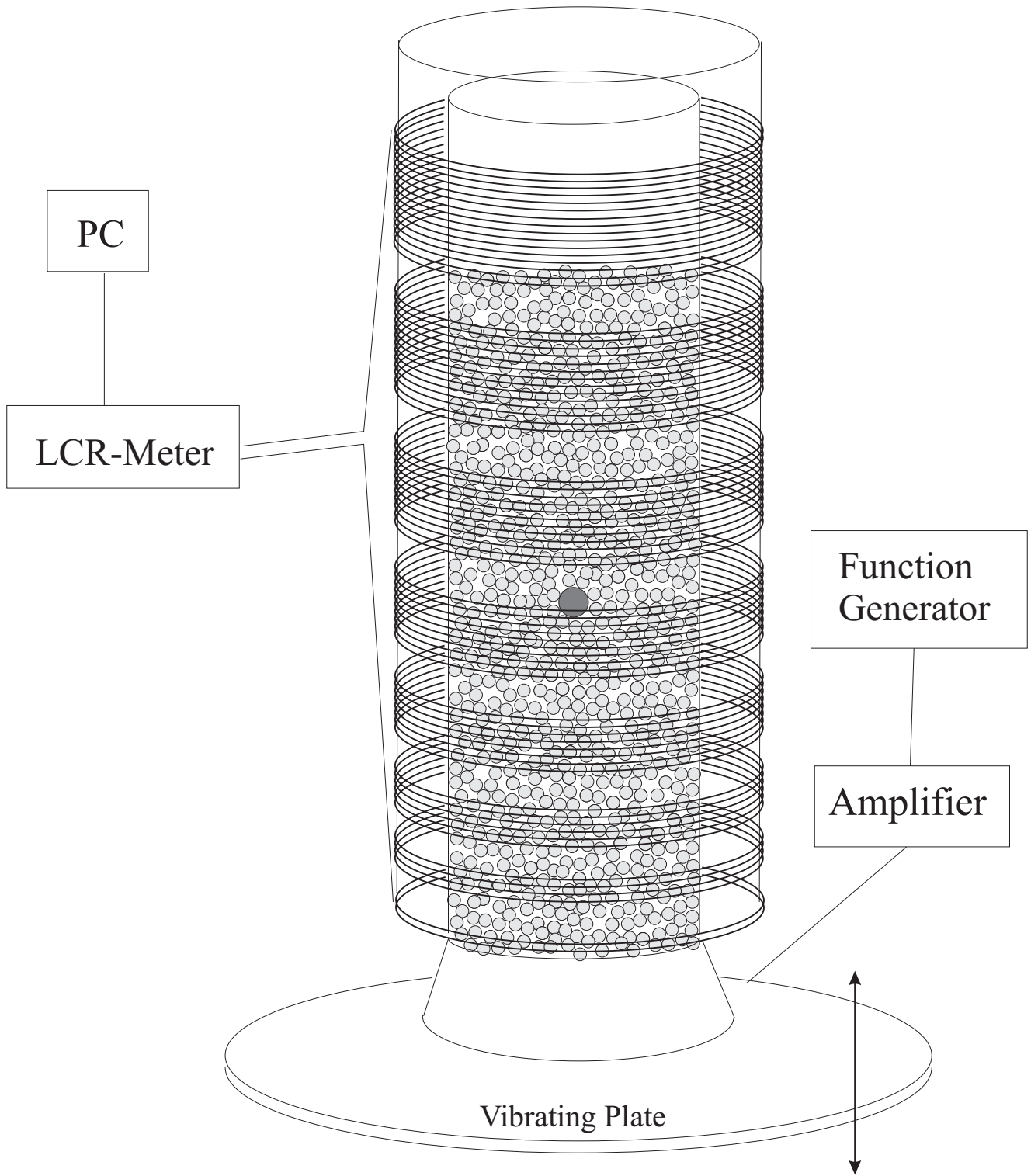
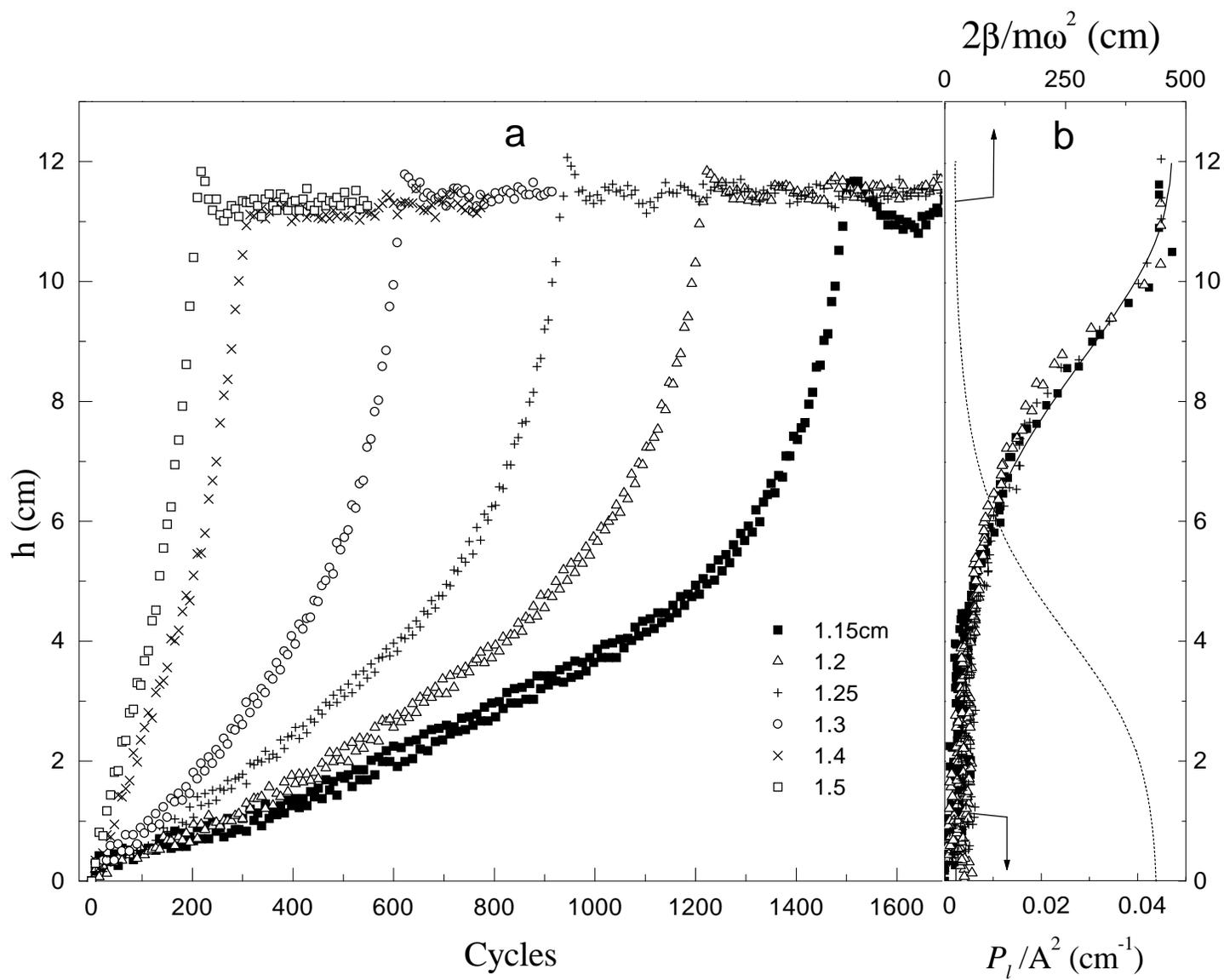
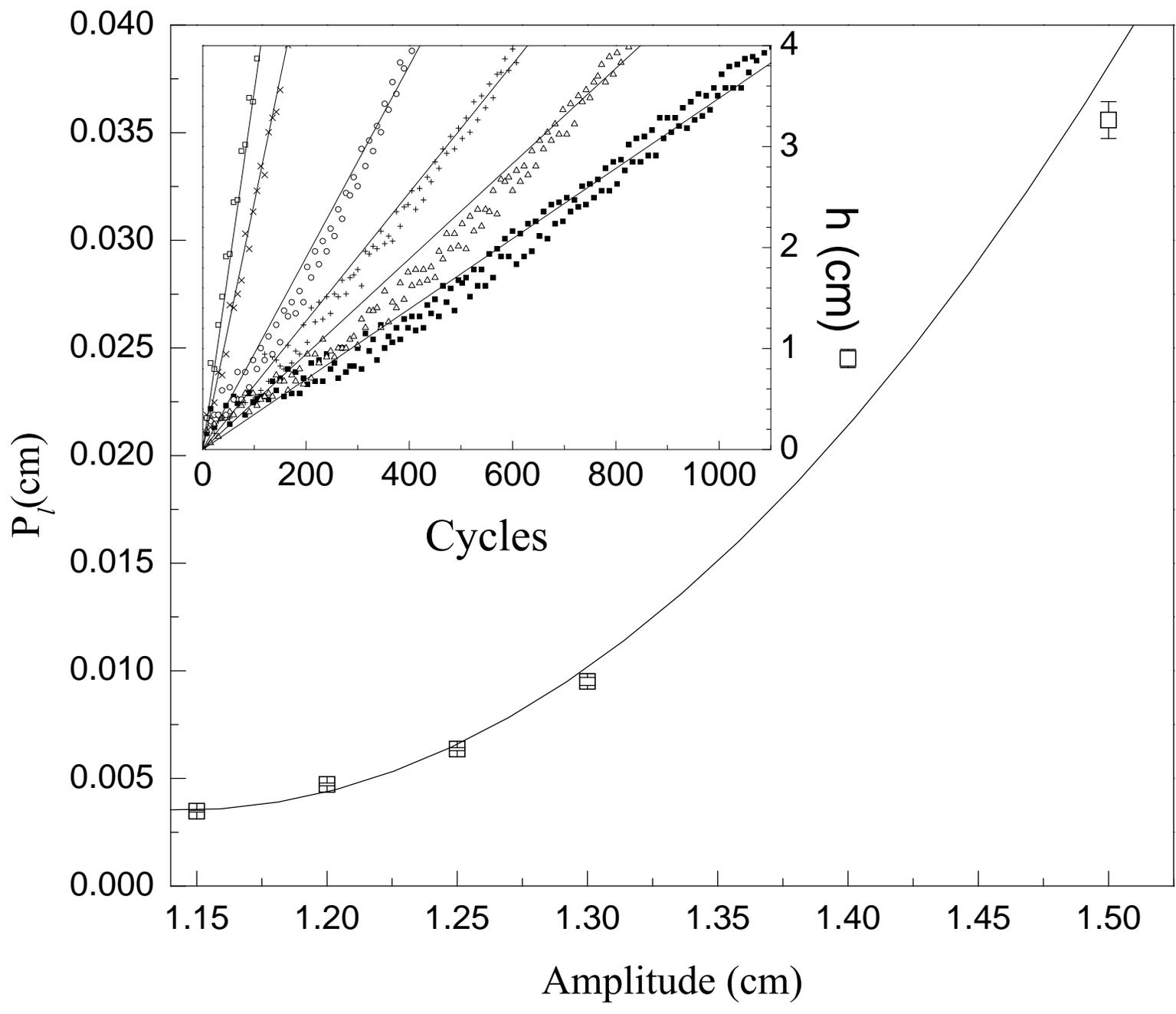
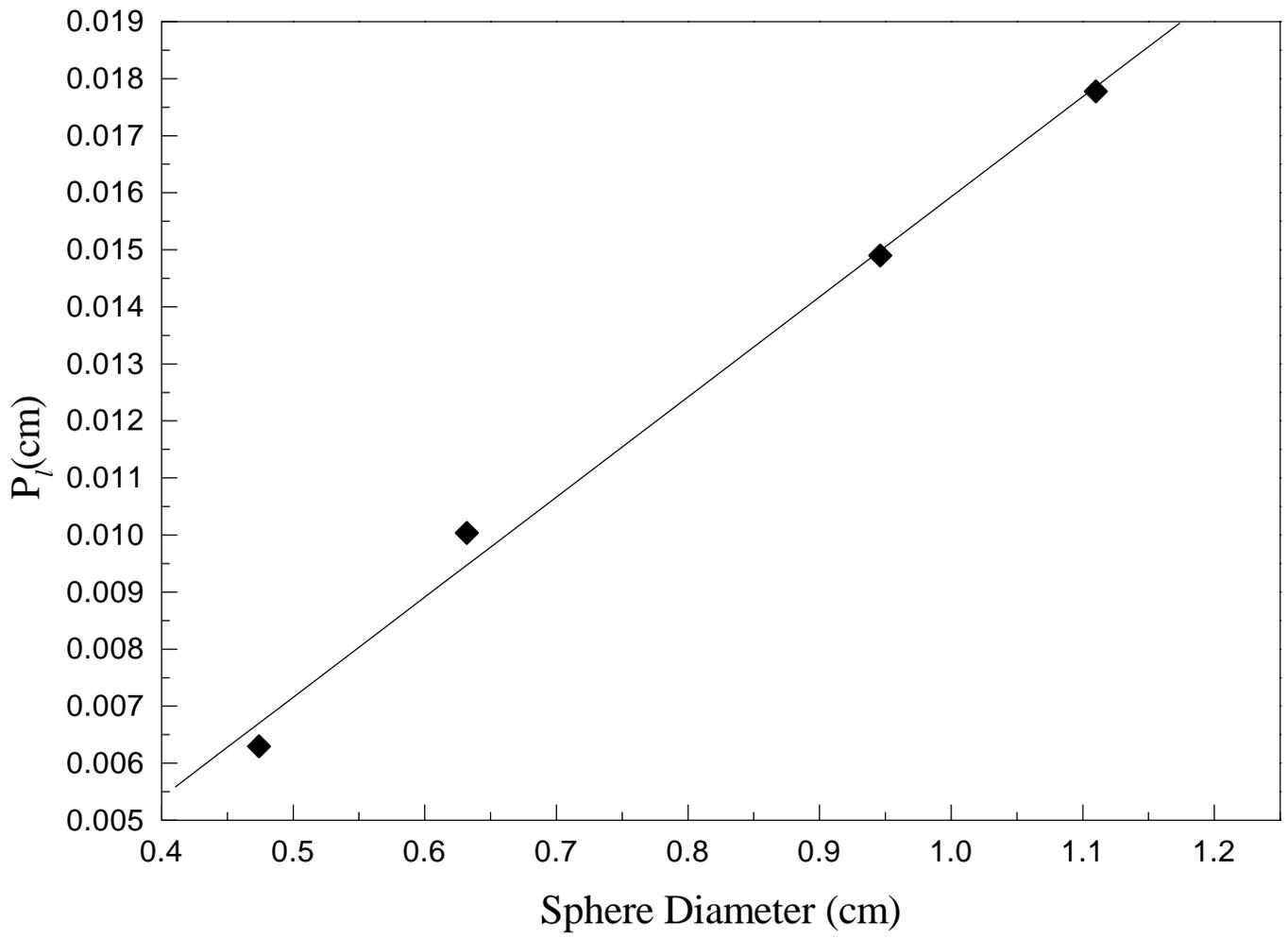
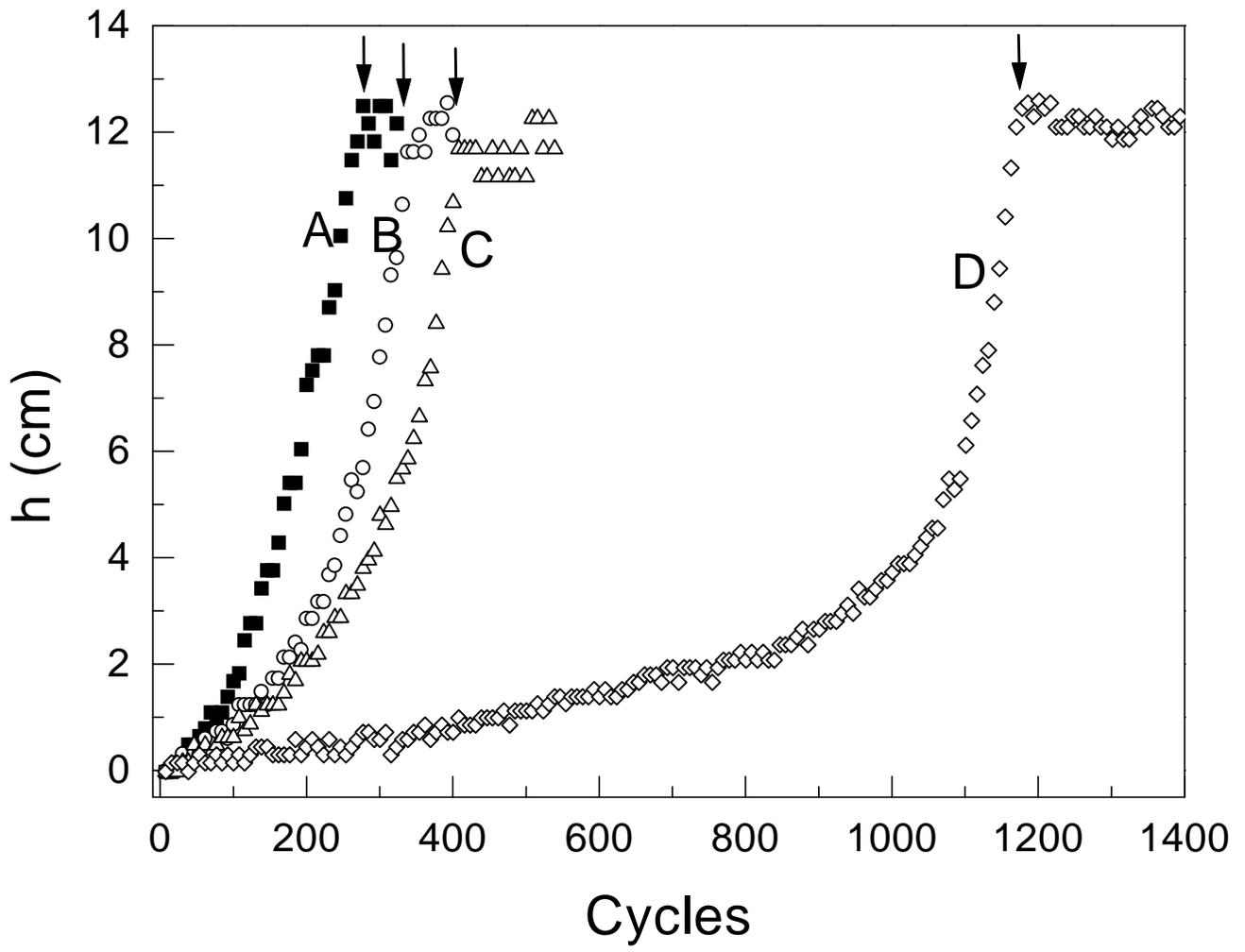
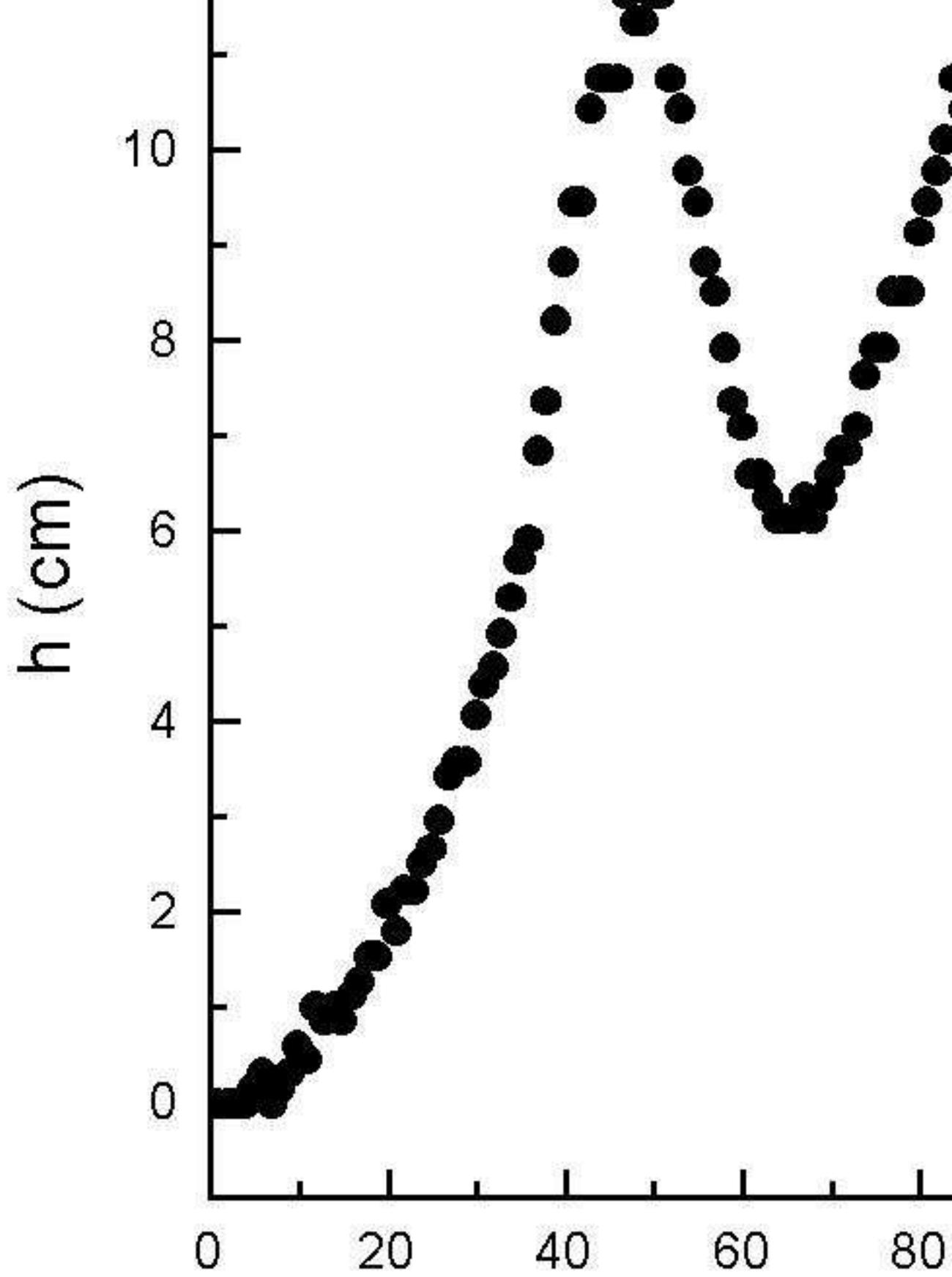
\end{document}